# Facets and Typed Relations as Tools for Reasoning Processes in Information Retrieval

Winfried Gödert

Cologne University of Applied Sciences, Institute of Information Science, Cologne, Germany

`winfried.goedert@fh-koeln.de`

**Abstract**. Faceted arrangement of entities and typed relations for representing different associations between the entities are established tools in knowledge representation. In this paper, a proposal is being discussed combining both tools to draw inferences along relational paths. This approach may yield new benefit for information retrieval processes, especially when modeled for heterogeneous environments in the Semantic Web. Faceted arrangement can be used as a selection tool for the semantic knowledge modeled within the knowledge representation. Typed relations between the entities of different facets can be used as restrictions for selecting them across the facets.

**Keywords**. Knowledge representation. Facets. Typed relations. Inferences. Information retrieval

## 1    Introduction

Knowledge representation regarded as information retrieval tool has to find solutions for representing knowledge elements beneficial for human use. Traditional indexing languages are strongly oriented towards a cognitive interpretation of their content representatives, the entities. This interpretation is primarily based on the user's knowledge and formally supported only by providing some different types of relationships between the entities. Normally, no formal definition of an entity is given, e.g. by assembling the set of content characterizing attributes. Assigned relationships between terms have to be verified mainly by plausibility. Likewise, no criteria are provided for the attributes that have been chosen for the decision to establish just these and no other relationships.

Due to these circumstances, the represented knowledge usually can only be retrieved based on the entities. Inference processes along the relational paths cannot be performed. Only forms of drill down searches supported by the hierarchical structure of the knowledge representation can sometimes be processed. The quality of the result sets strongly depends on the transitivity of the established hierarchies, a formal property that cannot always be taken for granted [3], [5], [18]. In this respect, indexing languages differ from knowledge structures built upon the principles of formal knowledge representations and that are seen as basis for semantic statements in Semantic Web environments.

We make a proposal for combining the two approaches in order to build retrieval systems with enhanced power. The aim is to discriminate a specific documents' *aboutness*[1] as far as its *a posteriori* content is regarded. Part of this proposal is the compatibility between a human understanding of the knowledge represented and to ensure the formal correctness of reasoning along the relational paths by machine inferences. The main methodological tools are faceted representation of entities, typed relations between the entities, and inferences along the relational paths.

We start the discussion by presenting a short summary of essentials necessarily to be observed for the task of indexing understood as a statement about a documents' content, its aboutness.

## 2      Aboutness and Indexing

Providing index terms can only be justified if the corresponding concepts are covered issues within the context of the document. Generally, this cannot be seen only with respect to isolated terms. Every concept is part of a semantic context and may be treated in syntactic connections to other concepts. The sum of all covered concepts and their embedding into semantic or syntactic relationships may be seen as the documents' aboutness. Indexing should represent this aboutness by means of a knowledge representation containing conceptual entities and semantic relationships between them. We will discuss the requirements to be observed when solving some non-trivial indexing and retrieval tasks.

The semantic context gives rise for the establishment of *a priori* relationships between the elements of an indexing language. Usually there are three types of relationships, *synonymy*, *hierarchy*, and *association*. The decision, which type of relationship should be established between two concepts is based on an intellectual content analysis of the concepts. Formal characteristics are usually not given, neither for the content definition of the elements nor for the relationships. Consequently, the decision whether an element should be used to represent a part of the documents' aboutness as an index term is also based on the elements' content by intellectual interpretation.

*A posteriori* relationships specify the context of the content elements on a second level. An identical sum of concepts may constitute a different aboutness if they are connected by different syntactical relationships. For example, one may represent the role 'to be an agent of an action' or 'to be the object of an action'. Up-to-now, there is a controversial discussion about the extent to which one should deal with this type of relationships. Their total amount is too large for including them all into indexing and retrieval environments. Furthermore, as precision tools they require additional effort.

Our concern is to discuss some characteristic features of the aboutness and its representation by the semantic features of an indexing language. We give some simple examples. Let us assume we are interested in *all* documents on *songbirds*. One may think that processing such a search only requires a single index term, e.g. *singing birds* to retrieve a complete result set. But there are documents about songbird species that have been indexed by a more specific index term, e.g. *titmice*. Ideally, the knowledge structure for the index terms represents such connections by hierarchical relationships. Automatically collecting the set of all entries along the hierarchical trail could help to generate the complete result set for *songbirds*. The success of such an approach depends on the inheritance of characteristics, in other words, the hierarchical relationship must fulfill the property of transitivity.

---

[1] We prefer to use the term aboutness instead of content, subject, or topic;
cf. http://www.iva.dk/bh/core concepts in lis/articles a-z/aboutness.htm

Our second example is described by topics that require more than one term as result of a form of coordinate indexing and as search terms for post-coordinate retrieval, e.g. combining the terms by Boolean operators or other syntactical devices. To give an impression, our aforementioned example could be enhanced to the topic *migratory behavior of songbirds*. This case would require that *migration behavior* as well as *singing birds* have been indexed and that they can be searched in combination.

Much more crucial is the type of topics represented by the third example, *songbirds with migratory instinct*. In this case, *migratory instinct* shall not be a covered topic, so it may not have been indexed for the documents to be found. *Migratory instinct* is only a constraining property for selecting the appropriate species of songbirds. There exist a lot of songbird species without such a behavior. Therefore, *migratory instinct* has to be modeled as a constraint for selecting the appropriate entities of the knowledge representation for subsequently generating the result set.

We can generalize our example to the key question: how should such constraints be represented in a knowledge representation in order to benefit indexing and to support inference mechanisms in retrieval environments? We will come back to the example and its generalization in Section 5.

## 3    Indexing and Facets

Basic tool for intellectual indexing is a structured collection of controlled terms. The concepts of such an indexing language are usually arranged according to their characteristics as determined by intellectual analysis. For building proper hierarchies over several stages of the relational paths, inheritance of these characteristics should be observed. This means that each hierarchically subordinated concept has all characteristics of the parent concept and at least one more. This requirement can best be met by avoiding poly-hierarchies or poly-dimensional connections. This especially means that the additional features must originate from a common aspect area or a certain categorical facet. If the generic context for the determination of characteristics is changed, poly-hierarchies become unavoidable and the result most commonly is a pre-combined conceptual ordering with insufficiently expressed criteria.

As an example one can consider the following terms: table, wooden table, glass table, kitchen table, living room table, art nouveau table, desk, changing table, side table. Any attempt to organize these concepts – or even an extension – into a single hierarchy must fail because of the reasons stated. Therefore, a subordination by changing aspects, as it is often found in pre-combined classifications or thesauri with extensive use of conceptual composition (e.g. by using compound nouns), cannot be a suitable condition for drawing conclusions by relational inferences.

To support inferences we need a faceted structure preserving the generic contexts within each hierarchy string. It is commonly agreed that a faceted arrangement of knowledge elements supports best the requirements for indexing and retrieval [2]. Using faceted indexing terms for representing the documents' aboutness allows to specify each content component at the desired level of granularity without ignoring the structural requirements of building transitive relational paths. Figure 1 may serve as a visualization of a faceted structure suitable for the topic: *Illustrated book of European songbirds*, composed by subjects of three aspects.

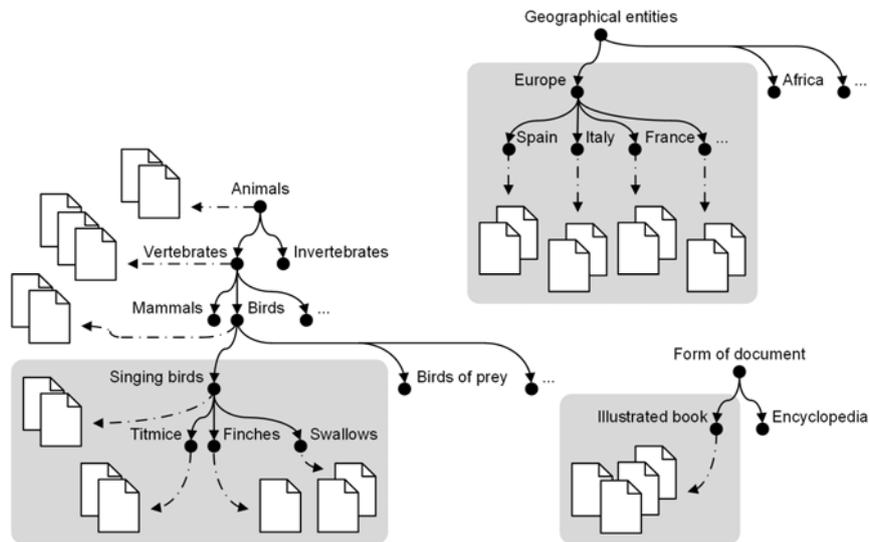

**Fig. 1.** Knowledge structure with facets, hierarchical inferences and associated documents

Such an arrangement has strong advantages for establishing proper hierarchies that avoid ambiguities. Connections established by common cultural use, scientific tradition, or statistical co-occurrences may be seen as beneficial but cannot be expressed by relationships within the facets. Integrating a topic like *European songbirds with migratory instinct* as a single entity would create a pre-combined knowledge structure with poly-dimensional connections between the entities. It cannot be properly connected to the entities of only one facet without violating the requirements of transitivity and transparency of the resulting structure. We have to find another solution to represent such topics by the entities and the structural features of a knowledge representation.

## 4 Facets and Typed Relations

The conceptual parts of our example can be differentiated into faceted components. One facet may be characterized as a taxonomical facet (*songbirds*) and a second one as a behavioral facet (*migratory instinct*). The connection between both facets can be represented by a relationship specified as a type of association from an inventory of typed relationships. The general pattern of such connections is given in Figure 2 with an exemplification of three types of associative relationships between the hierarchical organized entities of four facets.

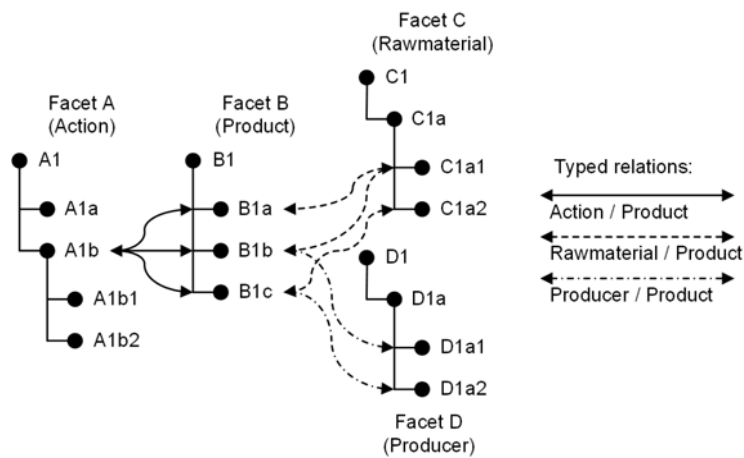

**Fig. 2.** Faceted systematic structure with typed relations between the facets

Analyses for developing such inventories with resulting proposals have been reported in the literature, e.g. in [12], [19]. Basis of these proposals were some theoretical studies reported for example in [8], [17]. We will use a condensed inventory [6, 7] for our discussion in Section 5. If characterized by transitivity properties it will become possible to draw inferences along the paths of typed associative relationships between entities of different facets and not only for entities connected by hierarchical relationships. In this way we want to demonstrate the potential of combining faceted structures with inferences along the paths of combinations of different typed relations.

## 5 Typed Relations and Inferences

Rule-based (formal) reasoning is supported by inferences about the underlying knowledge. For indexing and retrieval, this knowledge is given by the entities and the structure between them. Both elements, the indexed entities and also the relationships between them should therefore be used to form result sets.

Within our approach we distinguish two types of filter processes. One type allows drawing inferences on the *a priori* statements modeled in the knowledge representa-tion. The second type is applied to the results of an indexing process representing the *a posteriori* statements of a document. This second type is the more common one, typically realized by the use of Boolean operators or other syntactical devices.

Figure 3 gives a visualization of the requirements for the first process by using abstract labels. This illustration stresses the importance of structural properties of such knowledge representations instead of only interpreting the content of the semantic entities intellectually. The labels should be read as follows:

— $E_jF^i$ – Entity j in Facet i
— $E_jxF^i$ – Entity in Facet i that is in a *is a* relationship to $E_jF^i$, $x \in \{a\text{-}z\}$
— $E_jxyF^i$ – Entity in Facet i that is in a *is a* relationship to $E_jxF^i$, $x \in \{a\text{-}z\}$, $y \in \{a\text{-}z\}$
— $Rel_i$ - Relationship of type i

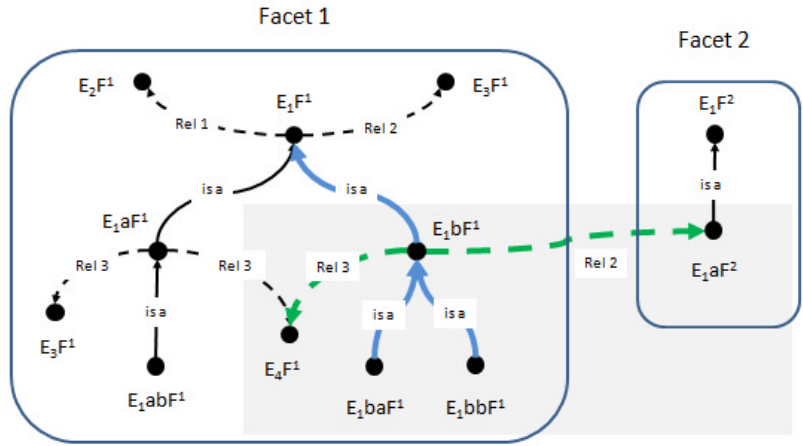

**Fig. 3.** Knowledge representation with typed property assignments and inferences

This visualization stands for a search interest that may require the selection of *all* entities that can be characterized as 'all entities of facet 1 that can be described as entities related to $E_1F^1$ by a *is a* relation under the restriction given by an entity $E_1aF^2$ of facet 2', in short $E_1b\$F^1 *_{Rel2}* E_1aF^2$. Neither the entity $E_1F^1$ nor the entity $E_1aF^2$ should be used as a search term, the search should only be processed by the entities $E_1b\$F^1$.

Considering *songbirds with migratory instinct* as an example of this abstract setting, neither *songbirds* nor *migratory instinct* must have been assigned as index terms to the relevant documents. We are looking for documents that make statements *about* certain concepts (*songbirds*) with specified properties (*have migratory instinct*) without these properties themselves being covered in the documents.

Facet 1 of our knowledge representation contains the knowledge about the entity $E_1F^1$ *(Singing birds)*, e.g. that $E_1bF^1$ is a songbird species. A typed relation (Rel2) to entity $E_1aF^2$ of facet 2 provides the constraint (*has migratory instinct*) for all entities $E_1b\$F^1$ of facet 1 that is not valid for $E_1F^1$. An inference about the hierarchical rela-tion paths allows to generate the set of all applicable songbirds and thus to carry out the respective search. The dashed line connecting $E_1bF^1$ to $E_1aF^2$ gives an illustration for this inference.

The example provides a proper illustration of the aforementioned abstract rule to choose the appropriate hierarchical node for linking it to another node of a second facet by a typed relation in order to avoid the necessity for exception clauses. For better comprehension we will illustrate this statement by discussing three figures that present different small knowledge structures. It is not difficult to generalize the examples' structural content to the general situation.

Figure 4 shows first an assignment of typed relations to an entity *Singing birds* that must be qualified as incorrect. One of these relationships is given by '*Singing birds* has *Migration pattern*'.

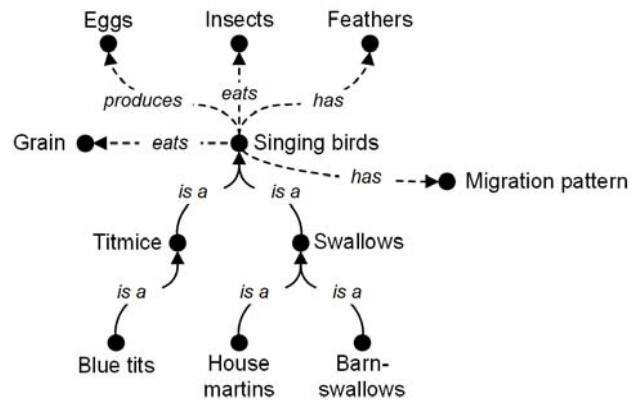

**Fig. 4.** Knowledge structure with incorrect connection of typed relations

Of course there are songbirds having a migratory instinct, but this is not true for *all* species. A solution for modeling this fact might be to connect only those subordinated species to *Migration pattern* that really do have a migratory instinct. Despite a lot of effort this solution causes a lack of transparency within the knowledge structure. This fact is indicated in Figure 5 for connecting only very few instances. It should be imagined as a simulation of the general situation with possibly hundreds of connections.

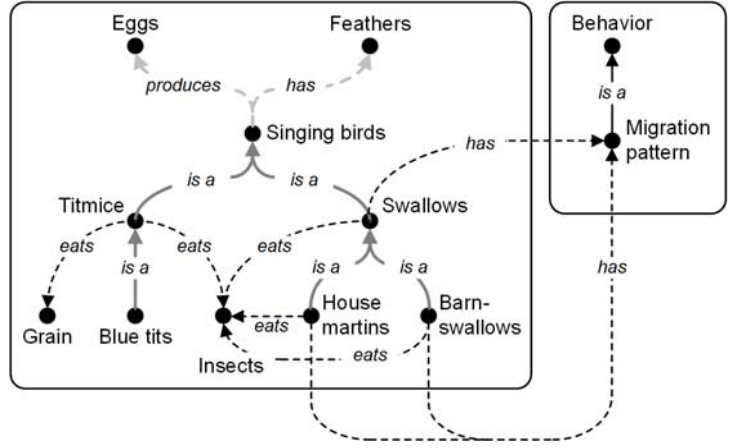

**Fig. 5.** Knowledge structure with typed property assignments but without inferences

Another solution of this problem may be the application of exception clauses following the approach of expert systems.

Our proposal given by Figure 3 combines mainly two elements,

– choosing the appropriate hierarchical node for linking it to another node by a typed relation,
– inferences along the relational paths.

Figure 6 gives an exemplification of the general setting for our example. Not all species of songbirds are connected to *Migration pattern* but only those with migratory instinct. To avoid the danger of building structures with less transparency and to ensure the content-orientated consistency of the formal properties, only the species of respective highest hierarchical level have been chosen for the connections.

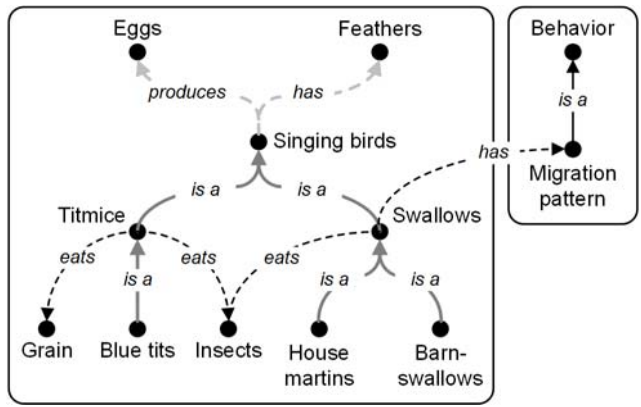

**Fig. 6.** Knowledge representation with typed property assignments and inferences

A further enhancement for using inferences in the domain of information retrieval is given by building result sets along relational paths of combinations of different relationships. As has been mentioned, such a mechanism is not well supported by traditional indexing languages. Therefore, it may be beneficial to analyze different combinations of semantic *a priori* relations in order to make some propositions about transitive inferences along the relational paths. Because of the content-based decisions about the type of relationships in indexing languages it is not possibly to give exact proofs for such transitivity statements. Especially for associative relationships one is restricted to plausibility arguments. We present a summary of our findings in three tables, where we make use of the following enhanced inventory of the usual relationships [6, 7].

    Equivalence
       Synonym
    Hierarchy
       Abstraction, generic context
       Whole / Part
       Abstraction, generic context
       Whole / Part
    Chronological context
       Earlier / Later
       Later / Earlier
       Earlier / Later
       Later / Earlier
    Association
       Unspecific association
       Raw material / product
       Causality (cause – effect)
       Person as actor / action
       Institution as actor / action
       Person as actor / product
       Institution as actor / product
       Action / product

The guiding principle for our transitivity statements is always the inheritance of characteristics. As has been discussed, for the hierarchical relationships this is supported by a faceted approach in order to avoid ambiguities. A more detailed analysis and presentation of the arguments has been given in [7].

The transitivity statements are always indicated in the last column by the symbols:

  +: Transitivity is given
  -: Transitivity cannot be expected
  O: Not allowed for indexing languages

Table 1 shows the transitivity statements for combining relationships of same type.

**Table 1.** Transitivity in case of same type of relationships

| Type of relation | Relation 1 | Relation 2 | Transitivity |
|---|---|---|---|
| Equivalence | Synonym | Synonym | O |
| Hierarchy | Abstraction, generic context | Abstraction, generic context | + |
| | Whole / Part | Whole / Part | + |
| | Abstraction, generic context | Whole / Part | - |
| | Whole / Part | Abstraction, generic context | - |
| Chronological context | Earlier / Later | Earlier / Later | + |
| | Later / Earlier | Later / Earlier | + |
| | Earlier / Later | Later / Earlier | - |
| | Later / Earlier | Earlier / Later | - |
| Association | Unspecific association | Unspecific association | - |
| | Raw material / product | Raw material / product | + |
| | Causality (cause – effect) | Causality (cause – effect) | + |
| | Person as actor / action | Person as actor / action | - |
| | Institution as actor / action | Institution as actor / action | - |
| | Person as actor / product | Person as actor / product | - |
| | Institution as actor / product | Institution as actor / product | - |
| | Action / product | Action / product | - |

Table 2 shows the transitivity statements for combining different types of hierarchical relationships.

**Table 2.** Transitivity in case of different types of hierarchical relationships

| Type of relation | Relation 1 | Relation 2 | Transitivity |
|---|---|---|---|
| Hierarchy | Synonym | Abstraction, generic context | + |
| | Synonym | Whole / Part | + |
| | Abstraction, generic context | Synonym | O |
| | Abstraction, generic context | Synonym | O |
| | Whole / Part | Synonym | O |
| | Whole / Part | Synonym | O |
| Chronological context | Synonym | Earlier / Later | + |
| | Synonym | Later / Earlier | + |
| | Earlier / Later | Synonym | O |
| | Later / Earlier | Synonym | O |
| | Abstraction, generic context | Earlier / Later | + |
| | Abstraction, generic context | Later / Earlier | + |
| | Earlier / Later | Abstraction, generic context | + |
| | Later / Earlier | Abstraction, generic context | + |
| | Whole / Part | Earlier / Later | + |

| | | |
|---|---|---|
| Whole / Part | Later / Earlier | + |
| Earlier / Later | Whole / Part | + |
| Later / Earlier | Whole / Part | + |

Finally, Table 3 shows the transitivity statements for combining different types of hierarchical relationships with typed associative relationships. Only combinations with a positive statement have been included.

**Table 3.** Transitivity for combinations of typed associative with hierarchical relationships

| Type of relation | Relation 1 | Relation 2 | Transitivity |
|---|---|---|---|
| Association | Unspecific association | Abstraction, generic context | + |
| | Unspecific association | Whole / Part | + |
| | Unspecific association | Earlier / Later* | + |
| | Raw material / product | Abstraction, generic context | + |
| | Raw material / product | Whole / Part | + |
| | Raw material / product | Earlier / Later* | + |
| | Action/ product | Abstraction, generic context | + |
| | Action/ product | Whole / Part | + |
| | Action/ product | Earlier / Later* | + |
| | Person as actor / action | Abstraction, generic context | + |
| | Person as actor / action | Whole / Part | + |
| | Person as actor / action | Earlier / Later* | + |
| | Institution as actor / action | Abstraction, generic context | + |
| | Institution as actor / action | Whole / Part | + |
| | Institution as actor / action | Earlier / Later* | + |
| | Causality (cause – effect) | Abstraction, generic context | + |
| | Causality (cause – effect) | Whole / Part | + |
| | Causality (cause – effect) | Earlier / Later* | + |
| | Person as actor / product | Abstraction, generic context | + |
| | Person as actor / product | Whole / Part | + |
| | Person as actor / product | Earlier / Later* | + |
| | Institution as actor / product | Abstraction, generic context | + |
| | Institution as actor / product | Whole / Part | + |
| | Institution as actor / product | Earlier / Later* | + |

* Also: Later / Earlier

Such a list of transitivity statements may be used for a formal mark-up in knowledge representations when characterizing the properties of combinations of typed relationships for the design of semantically enhanced search environments.

## 6   Web Retrieval and Semantic Interoperability: an Outlook

The ideas presented in this paper can be used in any indexing and retrieval environment. But they may be especially useful when applied in heterogeneous information environments, e.g. in the context of Semantic Web applications and its standards for semantic representation. To demonstrate this usefulness in principle, a prototype with a search form was built at the *Institute of Information Science* at the *Cologne University of Applied Sciences* and can be accessed via a Web interface[2]. This prototype was based on existing data originating from the database *Literatur zur Informationserschließung (Information science literature)*[3]. The scope of the documents corresponds to an extract from the *ASIS&T Thesaurus* [1], which was transformed into a *topic map* ([4], [10], [15, 16]) by using the software *Ontopia*[4]. This environment was chosen for its special potential to represent entities and the relationships between them by typed relations as topic maps. The features of *Ontopia* comprise a visualization of the topic map and an ability to support a Web retrieval system. We refrain from giving more details or examples in this paper, more material has been presented in [6, 7].

Web retrieval and heterogeneous information environments include issues of semantic interoperability. Compared with the already given suggestions in the *SKOS* recommendations [9], [13, 14] or in *ISO 25964* [11], using typed relations may provide an advanced support for implementing inference processes and methods of faceted retrieval across heterogeneous indexing languages. The semantic representation standards *RDF* or *OWL* support an enhancement of existing relationships without a need for complete reorganization of the indexing language.

Further research and practical implementation projects will be necessary to get benefit for real information environments constituting of data records indexed by one or more of the well-known knowledge organization systems. This paper could only present a sketch of the approach. Any realization in a real world setting requires the combination of already elaborated linked data techniques with a labor intensive restructuring of knowledge organization systems used in heterogeneous indexing and retrieval environments.

---

[2] http://ixtrieve.fh-koeln.de/ghn/
[3] http://ixtrieve.fh-koeln.de/LitIE/
[4] http://www.ontopia.net